\documentstyle[12pt,amstex,amssymb,righttag]{article}

\begin{document}
\renewcommand{\thefootnote}{\fnsymbol{footnote}}

\thispagestyle{empty}

\vspace*{-1cm}
\begin{center}
{\Large \bf Analysis of Maxwell Equations in a Gravitational Field}

\vspace{3mm}
by\\
\vspace{3mm}
{\sl Sortelano Araujo Diniz$^{\ast}$} and 
{\sl Carlos Pinheiro$^{\ast}$}

\vspace{3mm}
$^{\ast}$Universidade Federal do Esp\'{\i}rito Santo, UFES.\\
Centro de Ci\^encias Exatas\\
Av. Fernando Ferrari s/n$^{\underline{0}}$\\
Campus da Goiabeiras 29060-900 Vit\'oria ES -- Brazil\\

\noindent
{e-mail:diniz@@cce.ufes.br or fcpnunes@@cce.ufes.br}

and

{\sl F.C. Khanna}\\
\footnotesize{Theoretical Physics Institute, Dept. of Physics\\
University of Alberta,\\
Edmonton, AB T6G2J1, Canada\\
and\\
TRIUMF, 4004, Wesbrook Mall,\\
V6T2A3, Vancouver, BC, Canada.
khanna@@phys.ualberta.ca}
\end{center}

\vspace{3mm}
\begin{center}
Abstract
\end{center}
In a gravitational field, we analyze the Maxwell equations, the
correponding electromagnetic wave and continuity equations. A
particular solution for parellel electric and magnetic
fields in a gravitational background is presented. These solutions
also satisfy the free-wave equations and the phenomenology suggested
by plasma physics.

\newpage
\section{Introduction}
\setcounter{footnote}{0}
\paragraph*{}
Plasma and Astrophysical Plasma physicists support the possible existence of
electromagnetic stationary waves with parallel $\vec{E}$ and
$\vec{B}$ fields and consequently having a null Poynting
vector \cite{um,oito}. K.R. Brownstein \cite{um} points out that
these waves may emerge as solutions of the vector equation 
$\vec{\nabla}\times \vec{V}=k\vec{V}$ ($\vec{V}$ is a vector field
and $k$ is a positive constant). Brownstein considers
this equation for the vector potential $\vec{A}$ as
\begin{equation}
\vec{\nabla}\times \vec{A}=k\vec{A}\ ,
\end{equation}
with a particular solution
\begin{equation}
\vec{A}=a\left[i\sin kz +\hat{j}\cos kz \right]\cos \omega t
\end{equation}
and takes the associated electric and magnetic field as
\begin{equation}
\vec{E}=-\ \frac{1}{c}\ \frac{\partial \vec{A}}{\partial t}=ka
\left[i\sin kz+\hat{j}\cos ka\right]\sin \omega t \,
\end{equation}
and
\begin{equation}
\vec{B}=\vec{\nabla}\times \vec{A}=ka\left[\hat{i}sin kz+j\cos
kz\right]\cos \omega t\ .
\end{equation}
The Brownstein fields (1.3) and (1.4) satisfy Maxwell equations and the
usual vacuum free-wave equation. Moreover, fields, $\vec{E}$ and
$\vec{B}$ and the vector potential, $\vec{A}$, are parallel
everywhere. The associated electromagnetic wave has a null Poynting
vector and, so, does not propagate energy. The behavior of this wave
is just like the phenomenology suggested by Plasma and Astrophysical 
Plasma Physics.

But it is pointed out \cite{nove} that this model is not complete, at
least for Astrophysical Plasma. There it is argued that in an
Astrophysical Plasma, gravitation must be taken into account in a way
that the gravitational background can break the parellelism between the fields
$\vec{E}$ and $\vec{B}$. Consequently, the corresponding
eletromagnetic wave does not have a null Poynting vector. It is not a
stationary wave and propagates energy.

In this work, we consider the electromagnetic and gravitional
coupling and we analyze, in a gravitational background, the
corresponding Maxwell equations. We get the associated free-wave
equation for fields $\vec{E}$ and $\vec{B}$ and discuss the
corresponding electrostatic regime. That is, the situation in a
gravitational background in which the field $\vec{E}$ does not induce
the field $\vec{B}$ and vice-versa.

And we show that, in a particular situation, electric field $\vec{E}$
may depend on time but does not induce magnetic field $\vec{B}$ with
or without sources. Finally we show particular electrostatic
and magnetostatic solutions for Maxwell equations without sources in a
gravitational background. These solutions are such  that the electric
field, $\vec{E}_0$ and the magnetic field, $\vec{B}_0$, both satisfy
the corresponding free-wave equations. Moroever these fields 
behave  as the
phenomenology in Plasma Astrophysics has suggested, that is,
$\vec{E}_0$ and $\vec{B}_0$ are parallel fields. Thus, the associated
electromagnetic wave is stationary, has a null vector Poynting and
does not propagate energy.

\section{Electromagnetic and Gravitational Coupling}\setcounter{equation}{0}

\paragraph*{}
The action for the gravitaional and electromagnetic coupling is
written as 
\begin{equation}
{\cal S}=\int \sqrt{-g}\left(-\ \frac{1}{4}\ F_{\mu \nu}F^{\mu \nu}\right)
d^4s\ ,
\end{equation}
requiring stationary action $(\delta S=0)$ the corresponding
Maxwell inhomogeneous equations in a gravitational background are
\begin{equation}
{\cal D}_{\mu}F^{\mu \nu}={\cal J}^{\mu}\ ,
\end{equation}
where the covariant derivative above is given as:
\begin{equation}
{\cal D}_{\mu}F^{\mu \nu}=\partial_{\mu}F^{\mu \nu}+
\Gamma^{\mu \nu}_{\mu \lambda}+\Gamma^{\nu}_{\mu \lambda}\Gamma^{\mu \lambda}=
{\cal J}^{\nu} \ .
\end{equation}
The corresponding homogeneous Maxwell equations are
\begin{equation}
{\cal D}_{\mu}F_{\nu \rho}+{\cal D}_{\nu}F_{\rho \mu}+
{\cal D}_{\rho}F_{\mu \nu}=0\ .
\end{equation}
The connection terms of this equation cancel each other in such a
way that this equation is the usual homogeneous Maxwell equation
\begin{equation}
\partial_{\mu}F_{\nu \rho}+\partial_{\nu}F_{\rho \mu}+\partial_{\rho}
F_{\mu \nu}=0\ .
\end{equation}
Finally, to analyse these equations we adopt the F.R.W. cosmological metric
\begin{equation}
dS^2=dt^2-\left(a(t)\right)^2\left\{\left(1-Ar^2\right)^{-1}dr^2+r^2d\theta^2+
r^2\sin^2\theta d\varphi^2\right\}\ ,
\end{equation}
where the term $a(t)$ is the scale factor and the constant $A$ may
assume values $A=1,0,-1$. Each value represents the associated
curvature of F.R.W. spatial metric.

The second term $\Gamma^{\mu}_{\mu \lambda}$ of the covariant
derivative equation (2.3) can be written as
\begin{equation}
\Gamma^{\mu}_{\mu \lambda}=-\ \frac{\partial h}{\partial x^{\lambda}}
\end{equation}
where
\begin{equation}
h=\ln \sqrt{-\tilde{g}}
\end{equation}
and $\tilde{g}$ is the metric determinant. In terms of the electric
field, $E^i=F^{0i}$, and the magnetic field
$B^i=\varepsilon^{ijk}F_{jk}$, the Maxwell eq. (2.4) and (2.3)
are explicitly given by
\begin{equation}
\vec{\nabla}\cdot \vec{E}=\rho_{(\vec{x})}+\vec{\nabla}h\cdot
\vec{E}\ ,
\end{equation}
\begin{equation}
\vec{\nabla}\cdot \vec{B}=0\ ,
\end{equation}
\begin{equation}
\vec{\nabla}\times \vec{E}=-\ 
\frac{\partial \vec{B}}{\partial t}\ ,
\end{equation}
\begin{equation}
\vec{\nabla}\times \vec{B}=\vec{J}(\vec{x})+\frac{\partial \vec{E}}
{\partial t}-\frac{\partial h}{\partial t}\
\vec{E}+\vec{\nabla}h\times \vec{B}\ .
\end{equation}
Taking the divergence of equation (2.12) and using eq. (2.9) we get
the continuity equation in a gravitational background
\begin{equation}
\frac{\partial \rho}{\partial t}-\frac{\partial h}{\partial t} \ \rho
+ \vec{\nabla}\cdot \vec{J}-\vec{\nabla}h\cdot \vec{J}=0 \ .
\end{equation}
which is expressed in a covariant way as
\begin{equation}
{\cal D}_{\mu}J^{\mu}=0 \ ,
\end{equation}
where the covariant derivative is ${\cal
D}_{\mu}=\partial_{\mu}-\partial_{\mu}h$. 

Without electromagnetic sources $(\rho =0\ ; \ \vec{J}=0)$ the free
electromagnetic field equations in a gravitational background are:
\begin{equation}
\vec{\nabla}\cdot \vec{E}=\vec{\nabla}h\cdot \vec{E}\ ,
\end{equation}
\begin{equation}
\vec{\nabla} \cdot\vec{B}=0 \ ,
\end{equation}
\begin{equation}
\vec{\nabla}\times \vec{E}=-\frac{\partial \vec{B}}{\partial t}\ ,
\end{equation}
\begin{equation}
\vec{\nabla}\times \vec{B}=\frac{\partial \vec{E}}{\partial t}-
\frac{\partial h}{\partial t}\ \vec{E}+\vec{\nabla}h\times \vec{B}\ .
\end{equation}
Taking the curl of equation (2.17), using equations (2.15) and
(2.18) and, since $\vec{\nabla}h$ does not depend explicitly on time,
we get the free-wave equation in a gravitational background for the
electric field:
\begin{equation}
\nabla^2\vec{E}-\frac{\partial^2\vec{E}}{\partial t^2}=\vec{\nabla}
(\vec{\nabla}h\cdot \vec{E})-
\frac{\partial \left(\displaystyle{\frac{\partial h}{\partial t}}\vec{E}\right)}
{\partial t}-\vec{\nabla}h\times (\vec{\nabla}\times \vec{E})\ .
\end{equation}

Similarly, taking the curl of eq. (2.18), and using eq. (2.16) and, since 
$\displaystyle{\frac{\partial h}{\partial t}}$ does not depend
explicitly on the spatial coordinates, we get the free-wave equation
in a gravitational background for the magnetic field:
\begin{equation}
\nabla^2\vec{B}-\frac{\partial^2\vec{B}}{\partial t^2}=-\vec{\nabla}
\times (\vec{\nabla}h\times \vec{B})-\frac{\partial h}{\partial t}\ 
\frac{\partial\vec{B}}{\partial t}\ .
\end{equation}

\section{Analysis of Maxwell Equations in a Gravitational
Background}\setcounter{equation}{0} 

\paragraph*{}
Now we  analyze the necessary conditions to obtain
electrostatic and magnetostatic solutions in the gravitational
background, that is, the conditions that the electric field $\vec{E}$
and the magnetic field $\vec{B}$ must satisfy in order that the
electric field $\vec{E}$ does not induce the magnetic field $\vec{B}$
and vice-versa.

>From eq. (2.11) it is clear that the gravitational field does not modify the
Faraday's law and so it is simple to conclude that the magnetic field
$\vec{B}$ must be stationary in order not to induce an electric field
$\vec{E}$. On the other hand, the Amp\`ere Law (2.12) has been
modified by gravitation. Thus, it is simple to see that even if the electric
field is stationary $\left(\displaystyle{\frac{\partial
\vec{E}}{\partial t}}=0\right)$ it may induce a magnetic field
$\vec{B}$ from the gravitation term 
$-\displaystyle{\frac{\partial h}{\partial t}}\ \vec{E}$. For
example, if we have a non-null stationary electric field 
$\left(\displaystyle{\frac{\partial \vec{E}}{\partial t}}=0\ ; \
\vec{E}\neq 0\right)$ and no current density $(\vec{j}=0)$, eq. (2.18)
becomes $\vec{\nabla}\times \vec{B}=-\ \displaystyle{\frac{\partial
h}{\partial t}}\ \vec{E}+\vec{\nabla}h\times \vec{B}$ and so it
implies that we necessarily have a non-null magnetic field $\vec{B}$,
since $\vec{B}=0$ is not solution for this equation. From eq.
(2.12) it is clear that even if the electric field $\vec{E}$
depends on time it shall not induce a magnetic field
$\vec{B}$ provided the condition below is satisfied:
\begin{equation}
\frac{\partial \vec{E}}{\partial t}-\ \frac{\partial h}{\partial t}\ 
\vec{E}=0\ .
\end{equation}
>From condition (3.1) and from eq. (2.9) it is simple to verify that
the electric field $\vec{E}(\vec{r},t)$ and the charge density $\rho
(t)$ can be written as:
\begin{equation}
\vec{E}(\vec{r},t)=(a(t))^3\vec{E}_0(\vec{r})\ ,
\end{equation}
and
\begin{equation}
\rho (\vec{r},t)=(a(t)^3)\rho_0(\vec{r})\ ;
\end{equation}
where $\vec{E}_0(\vec{r})$ is a stationary vector field and
$\rho_0(\vec{r})$ is stationary ``charge density''. Furthermore the 
current density $\vec{J}$ must satisfy the equation:
\begin{equation}
\vec{\nabla}\cdot \vec{J}-\vec{\nabla}h\cdot \vec{J}=0
\end{equation}
If the conditions (3.1)-(3.4) are verified and if the magnetic field
$\vec{B}$ is static, the electrostatic and magnetostatic
field-equations become
\begin{equation}
\vec{\nabla}\cdot \vec{B}=0 \ ,
\end{equation}
\begin{equation}
\vec{\nabla}\times \vec{B}=\vec{j}+\vec{\nabla}h\times \vec{B}\ ,
\end{equation}
\begin{equation}
\vec{\nabla}\cdot \vec{E}_0=\rho_0(\vec{r})+\vec{\nabla}h\cdot 
\vec{E}_0\ ,
\end{equation}
\begin{equation}
\vec{\nabla}\times \vec{E}_0=0 \ .
\end{equation}

Any solution of equation (3.4)-(3.6) for the charge current density
$\vec{J}(\vec{r},t)$ and for the magnetic field, $B(\vec{r},t)$, can
be combined with any solution of equations (3.2), (3.3), (3.7) and
(3.8). So under these conditions the magnetic field, $B(\vec{r},t)$,
does not induce the electric field, $E(\vec{r},t)$, and vice-versa.
This way we call these equations magnetostatic and electrostatic, but
we must remember the electric field is
not static but depends on time according to eq. (3.2).

\section{Solutions for Maxwell Equations without Sources in a
Gravitational Scenery}\setcounter{equation}{0}

\paragraph*{}

We now consider the Maxwell equations (2.9)-(2.12) without sources
$(\rho =0\ ; \ \vec{J}=0)$ with the condition that the electric field
$\vec{E}(\vec{r},t)$ satisfies eq. (3.1) and that the magnetic field
$\vec{B}$ does not depend explicitly on time. The Maxwell equations
for the electrostatic and magnetostatic field-equations without
sources in a gravitational field (3.5)-(3.8) are given by:
\begin{equation}
\vec{\nabla}\cdot \vec{B}=0 \ ,
\end{equation}
\begin{equation}
\vec{\nabla}\times \vec{B}=\vec{\nabla}h\times \vec{B}\ ,
\end{equation}
\begin{equation}
\vec{\nabla}\cdot \vec{E}_0=\vec{\nabla}h\cdot \vec{E}_0 \ ,
\end{equation}
\begin{equation}
\vec{\nabla}\times \vec{E}_0=0 \ .
\end{equation}

A particular solution in spherical coordinates is
\begin{equation}
E_0(\vec{r},t)=\frac{\alpha}{r\ \sin \theta}\Longrightarrow \vec{E}(\vec{r},t)=
\alpha \ \frac{(a(t))^3}{r\sin \theta}\ \hat{\varphi}
\end{equation}
and 
\begin{equation}
\vec{B}(\vec{r},t)=\beta \ \frac{r}{\sqrt{1-Ar^2}}\ \hat{\varphi}
\end{equation}
where $\alpha$ and $\beta$ are constants and $\hat{\varphi}$ is the
unit azimutal vector: $\hat{\varphi}=(-\hat{i}\sin \hat{\varphi}
+\vec{j}\cos \varphi )$.

A simple substitution show that these fields satisfy all the Maxwell
equations without sources (2.15)-(2.18). It is interesting to point
out that the electric and magnetic fields are parallel and satisfy
the free-wave eq. (2.19) and eq. (2.20). The electromagnetic wave has
parallel electric and magnetic fields, has a null Poynting vector, it
is a stationary wave and it does not propagate energy as suggested by
the Astrophysical Plasma phenomenology.

The term $\vec{\nabla}h$ and the unitary azimutal vector
$\hat{\varphi}$ are perpendicular vectors, so
$(\hat{\varphi},\vec{\nabla}h,\hat{\varphi}\times \vec{\nabla}h)$
form a complete set and any vector can be written as:
\begin{equation}
\vec{V}=V_{\varphi}\hat{\varphi}+V_h\vec{\nabla}h+V_{\varphi h}
(\hat{\varphi}\times \vec{\nabla}h)
\end{equation}

The vector components $V_{\varphi}$, $V_h$ and $V_{\varphi h}$ may
depend on time and spatial coordinates. General solutions $E(\vec{r},t)$
and $\vec{B}(\vec{r},t)$ for Maxwell equations in a gravitational
background using the ansatz (4.7) are being considered.

We conclude that in astrophysical plasma it is important to consider
electromagnetic and gravitational coupling and this coupling modifies
the free-wave equations for electric and magnetic fields. 

\subsection*{Acknowledgements:}

\paragraph*{}

The authors acknowledge J.A. Hela\"{y}el-Neto for discussions and
comments. Thanks are also due to G.O. Pires Pires for reading the
manuscript. Carlos Pinheiro is partially supported by the Conselho
Nacional de Pesquisa -- CNPq/Brazil.

\end{document}